\documentstyle{article}

\font\tenrm=cmr10
\font\tenit=cmti10
\font\elevenbf=cmbx10 scaled\magstep 1
\font\elevenrm=cmr10 scaled\magstep 1
\font\elevenit=cmti10 scaled\magstep 1

\font\ninerm=cmr9

\renewenvironment{thebibliography}[1]
 { \elevenrm
   \begin{list}{\arabic{enumi}.}
    {\usecounter{enumi} \setlength{\parsep}{0pt}
     \setlength{\itemsep}{3pt} \settowidth{\labelwidth}{#1.}
     \sloppy
    }}{\end{list}}

\def\la{\mathrel{\mathpalette\fun <}}
\def\ga{\mathrel{\mathpalette\fun >}}
\def\fun#1#2{\lower3.6pt\vbox{\baselineskip0pt\lineskip.9pt
    \ialign{$\mathsurround=0pt#1\hfill##\hfil$\crcr#2\crcr\sim\crcr}}}
\def\iitt{\elevenit}
\def\bbff{\elevenbf}
\def\mol{ Mo\hskip-4.5pt/\hskip1pt l}
\def\NPB#1#2#3{{\iitt Nucl. Phys.} {\bbff B#1} (19#2) #3}
\def\PLB#1#2#3{{\iitt Phys. Lett.} {\bbff B#1} (19#2) #3}
\def\PRD#1#2#3{{\iitt Phys. Rev.} {\bbff D#1} (19#2) #3}
\def\PRL#1#2#3{{\iitt Phys. Rev. Lett.} {\bbff #1} (19#2) #3}

\begin{document}
\begin{flushright}
CTP-TAMU-51/93
\end{flushright}
\begin{flushright}
ACT-19/93
\end{flushright}

\begin{center}
\vglue 0.6cm
{{\elevenbf ASPECTS OF NEUTRALINO DARK MATTER\footnote{
\ninerm\baselineskip=11pt Talk given at the International
Workshop: ``Recent Advances in the Superworld'', Houston
Advanced Research Center, The Woodlands, TX, April 14-16,
1993.}
\\}
\vglue 5pt
\vglue 1.0cm
{\tenrm KAJIA YUAN \\}
\baselineskip=13pt
{\tenit Center for Theoretical Physics, Department of Physics,
Texas A\&M University \\}
\baselineskip=12pt
{\tenit College Station, TX 77843-4242, USA\\}
\baselineskip=13pt
{\tenit Astroparticle Physics Group, Houston Advanced Research
Center (HARC) \\}
\baselineskip=12pt
{\tenit The Woodlands, TX 77381, USA\\}}

\vglue 0.8cm
{\tenrm ABSTRACT}

\end{center}

\vglue 0.3cm
{\rightskip=3pc
\leftskip=3pc
\tenrm\baselineskip=12pt
%\noindent
The possible solution of dark matter problem with neutralinos of
supersymmetric models within the supergravity framework is reviewed.
A novel correlation between the neutralino relic abundance
$\Omega_\chi$ and the soft supersymmetry breaking patterns
is demonstrated. It is explained that, this generic result
together with the proton-decay constraint could significantly reduce
the allowed parameter space of the minimal $SU(5)$ supergravity
model, and therefore makes this model more easily testable.
The prospect of obtaining further cosmological constraints from
underground experiments for the minimal $SU(5)$ supergravity model
is also briefly discussed.}

\vglue 0.6cm
{\elevenbf\noindent 1. Introduction}
\vglue 0.4cm
\baselineskip=14pt
\elevenrm
Low-energy supersymmetry is now widely believed to be one of the
most appealing ideas likely responsible for new physics beyond the
Standard Model \cite{Habertalk}, as such, it has a very good chance
to reveal itself at next generation of $pp$ or $e^+e^-$
colliders, one way or another \cite{Bargertalk}.
If we indeed live in a ``SuperWorld'', whatever that might be,
then low-energy supersymmetry must also have cosmological
consequences. This motivated the study of ``SuperCosmology'',
of which many exciting topics have been covered in previous
talk by Prof. Schramm \cite{Schtalk}.
In this talk, I will confine myself to the proposal
\cite{Wein,EHNOS} of solving the dark matter problem with
neutralinos of supersymmetric models, as one typical example
which ties up low-energy supersymmetry with cosmology.

Let me begin with some remarks about the so-called dark matter problem.
First of all, it is a problem caused by the huge discrepancy
between the amount of directly observed visible matter and
the total amount of matter in the Universe indirectly inferred
based on general relativity (mostly Newtonian dynamics) \cite{DMrv}.
One should keep an open mind that, if at large scale the law of
gravity is in fact somehow modified,
the dark matter problem itself could cease to exist \cite{Milg}.
I will not entertain such a possibility here, but instead just
follow the majority point of view.
In terms of the present-day cosmological density parameter
$\Omega\equiv\rho_0/{\rho_{\rm crit}}$, the visible matter
of the Universe only amounts to $\Omega_{\rm vis} \la 0.01$.
Big-bang nucleosynthesis indicates that the fraction
of critical density contributed by ordinary baryons is
$0.011 \la \Omega_B \la 0.12$. However,
current theoretical prejudice prefers a flat universe today,
in particular, inflation predicts $\Omega =1$ to a very
high precision. Furthermore, without dark matter as an
essential ingredient, the mechanism of structure formation
in the Universe, based on the gravitational growth of primeval
density inhomogeneities (Jeans instability), does not seem to
work for galaxy formation. There are also observational
evidences for dark matter coming from various measurements of
the present-day mass density of the Universe, performed at
different distance scales, among them the flat rotation curves
of spiral galaxies appears to be most compelling.
Based on these arguments, one concludes that dark matter must
exist and be mostly in non-baryonic form.

Accepting the dark matter problem as a fact,
now the question is: what is dark matter made of?
So far, we do not have a definite answer, but we do
have many speculations (maybe too many!). Instead of going
through the list of all proposed dark matter candidates,
which can be found elsewhere \cite{DMrv},
I will only consider one type of hypothetical particle dark
matter which is closely associated with low-energy supersymmetry,
{\elevenit i.e.}, the neutralino dark matter \cite{Wein,EHNOS}.
It should be noted that, from particle physics point of view,
massive neutrinos with mass around $30$ eV and axions are
also particularly attractive dark matter candidates in their
own right.

We know that low-energy supersymmetry predicts many new
particles yet to be discovered, it would be extremely
fortunate if some of these particles in fact constitute
the dark matter out there in the Universe.
To be a possible dark matter candidate, a supersymmetric particle
has to satisfy at least three requirements: (a)
to be stable or at least have a life-time comparable with the
age of the Universe so that it can be around today as a
cosmological relic; (b) to have only gravitational and weak
interactions; (c) to have right properties such that its
relic abundance comes close to the critical value $\Omega=1$.
In most supersymmetric models, to eliminate the disastrous
explicit baryon and lepton number violating interaction
terms which lead to proton decay at unacceptable level,
a discrete symmetry known as $R$-parity is often invoked,
such that all ordinary particles are $R$-even, but all
superparticles are $R$-odd. Therefore, if $R$-parity is
conserved, superparticles always couple to the ordinary
particles in pair, and as a result the lightest superparticle
(LSP) is stable. The LSP is also likely to be colorless and
electrically neutral, since otherwise it would have formed
heavy anomalous isotope with ordinary matter which would
have been observed already \cite{WSB,EHNOS}. These considerations
tell us that, with unbroken $R$-parity, the LSP in fact
could make a viable dark matter candidate, if requirement
(c) is also satisfied. Unfortunately, no general remarks can be made
about this requirement for LSP, mainly because we do not really know
what is the LSP. The lightest neutralino $\chi$, a mixture of neutral
gauginos and higgsinos, which I will simply call neutralino
in the rest of this talk, currently stands as a front runner for
the LSP. The basic argument in favor of this theoretical prejudice
is essentially that, so far there have been no cosmological,
experimental or theoretical reasons against it.
In any case, as a working hypothesis, I will assume that
the lightest nuetralino is indeed the LSP. As we will see later,
in the supergravity models, since the complete particle
mass spectrum of the model can be specified rather economically
in terms of only a few parameters, it becomes possible to
consistently check the validity of this very assumption.

In the following, I will discuss some recent work regarding
the prospects for neutralino dark matter within the supergravity
framework, paying special attention to the correlation between
the neutralino relic abundance $\Omega_\chi$ and the supersymmetry
breaking patterns. Taking the minimal $SU(5)$ supergravity model
as an explicit example, in which the physics at grand unification
scale come into play, I will demonstrate the implications of this
novel correlation, and in passing also address the technical issue
of appropriately treating the thermal average cross section.
In addition, assuming that neutralinos constitute the dark matter in the
Galactic halo, and using the upward-going muon events in
underground detectors as a possible signal from such neutralinos
captured in the Earth and the Sun, I will also briefly discuss
the prospect of further exploring the minimal $SU(5)$ supergravity
model.

\vglue 0.6cm
{\elevenbf\noindent 2. Neutralino Relic Abundance}
\vglue 0.4cm
Now let me examine the requirement (c) for neutralinos in
more detail. Will neutralinos provide just right amount of mass
to close the Universe? From earlier work \cite{Wein,EHNOS,Group1,Group2},
we have learned that the answer to this question is YES and NO.
Yes, because it is possible to find regions in the parameter space of
the supersymmetric models where neutralino relic abundance is just
what we would like it to be ($\Omega_\chi\sim 1.0$);
No, because it is equally true that there are regions where
either $\Omega_\chi \ll 1.0$ or $\Omega_\chi \ga 1.0$.
Needless to say, regions with $\Omega_\chi\sim 1.0$ are
cosmologically favored. The regions of $\Omega_\chi \ga 1.0$ should
be excluded on cosmological grounds, since where the Universe is
uncomfortably younger than about 10 billion years, and
this argument alone often leads to interesting constraints
on the models under consideration. On the other hand,
if $\Omega_\chi \ll 1.0$, the best one can learn is that
now neutralinos can not be the sole source of dark matter.
Despite this, I want to stress that models with too small values of
$\Omega_\chi$ are perfectly healthy, while on the contrary models
which predict $\Omega_\chi \ga 1.0$ are definitely in trouble.
Clearly, one would like to be able to identify these three
distinct cases as precisely as possible.

\vglue 0.2cm
{\elevenit\noindent 2.1 Calculation Procedures}
\vglue 0.1cm
The basic physics involved in calculating relic abundance
for any massive stable particles was well understood
some time ago \cite{LeeWein}, which can be applied to the
neutralino case. At very early time the Universe was
radiation dominated, all particles would be in thermal equilibrium,
neutralinos annihilate into other particle species, and
vice versa. As the temperature drops down below the
neutralino mass $m_\chi$, the annihilation process
becomes dominant and the neutralino number begins to decrease
due to Boltzmann suppression, until the interactions between
neutralinos ``freeze out'', which happens when the annihilation
is no longer able to keep pace with the expansion of the Universe.
After ``freeze out'', the number of neutralinos essentially
remains constant and the number density only reduces as
a consequence of the cosmological expansion.
All of these are neatly
embodied into the Boltzmann equation \cite{LeeWein}
\begin{equation}
{dn\over dt}=-3Hn-\langle{\sigma v_{\mol}}\rangle (n^2-n^2_0),
\label{eq:LW}
\end{equation}
where $n$ is the actual number density of the neutralinos,
$n_0$ is the density they would have in thermal equilibrium
at temperature $T$, $H=(dR/dt)/R$ is the Hubble expansion
parameter, and $\langle{\sigma v_{\mol}}\rangle$ is the
thermal-averaged product of annihilation cross section
and the M\o ller velocity of the annihilating neutralinos
in the cosmic comoving frame \cite{GG91}.

In practice, it is convenient to replace time $t$ in Eq.~(\ref{eq:LW}) by
the photon temperature $T$, this is because the present-day cosmic
background radiation (CBR) temperature can be measured rather accurately,
while determining the age of the Universe is a quite
different story (see e.g. Ref.\ \cite{Turner}). Using the
conservation of entropy, one can recast Eq.~(\ref{eq:LW})
into an convenient form involving $T$. Although there exist slightly
different approaches in the literature, I will closely follow
Ref.\ \cite{Watkins} here. The ``new'' Boltzmann equation now reads
\begin{equation}
{dq\over dx}=\lambda(x)(q^2-q^2_0)(x),
\label{eq:Blz}
\end{equation}
with
\begin{equation}
\lambda(x)=({4\over 45}\pi^3G_N)^{-1/2}{m_\chi\over \sqrt{g(T)}}
[h(T)+{1\over 3}m_\chi xh'(T)]\langle{\sigma v_{\mol}}\rangle ,
\label{eq:lbd}
\end{equation}
where $q\equiv n/(T^3h(T)),\,q_0\equiv n_0/(T^3h(T))$,
$x\equiv T/m_\chi$, and $g(T),\,h(T)$ $(h'(T)=dh/dT)$ are the
effective degrees of freedom associated with energy and entropy
density respectively \cite{Watkins}. Finally, the relic abundance
is given by
\begin{equation}
\Omega_\chi h^2_0 = 1.555 \times 10^8 (m_\chi/{\elevenrm GeV})h(0)q(0),
\label{eq:relic}
\end{equation}
where $h_0=H_0/(100 {\rm km} {\rm sec}^{-1} {\rm Mpc}^{-1})$
parameterize our ignorance
of the present-day value of the Hubble parameter ($0.5 \la h_0 \la 1.0$);
$h(0),\,q(0)$ are the present-day values of $h(T),\,q(T)$ respectively.

{}From Eqs.~(\ref{eq:Blz})-(\ref{eq:relic}) it is clear that
the calculation of relic abundance
typically involves three procedures:
(I) Computing $h(T),\,g(T)$,
in particular, $h(0)$; (II) Evaluating $\langle{\sigma v_{\mol}}\rangle$;
(III) Solving Eq.~(\ref{eq:Blz}) for $q(x)$ and find $q(0)$.
Again, one can often find different ways of implementing these
procedures in literature. For a detailed discussion
of our approach to (I) and (III), see the Appendix of
Ref.\ \cite{LNY}. It is worthy of mentioning that,
in solving Eq.~(\ref{eq:Blz}), a WKB approximation \cite{BBF} was
used in Ref.\ \cite{LNY}, which matches the solution $q(x)$ with $q_0(x)$
and provides the initial condition for Eq.~({\ref{eq:Blz}) at a point $x_0$
where the WKB solution fails. Although, it is the same in spirit to determine
$x_0$ here as to determine the so-called freeze-out temperature
in the standard Lee-Weinberg method \cite{Wein,EHNOS}, however,
one advantage of this new approach is that the accuracy
of the solution can be easily controlled as one desires.

The most accurate treatment so far available for the thermal average factor
$\langle{\sigma v_{\mol}}\rangle$ was given in Ref.\ \cite{GG91}.
In this remarkable paper, the authors were able to reduce
the multiple integrals associated with the thermal average
into a single integral, which can be rewritten in the
following convenient Lorentz invariant from \cite{LNY93}
\begin{equation}
\langle{\sigma v_{\mol}}\rangle ={1\over 4m^5_\chi x K^2_2(x^{-1})}
\int_{4m^2_\chi}^\infty ds(s-4m^2_\chi)^{1/2}K_1(\sqrt{s}/xm_\chi)\,w(s),
\label{eq:GG}
\end{equation}
in terms of the Lorentz invariant function \cite{Watkins}
\begin{equation}
w(s)={1\over 4}\int d\,LPS\,|{\cal A}(\chi\chi\to all)|^2.
\label{eq:wfuc}
\end{equation}
In Eq.~(\ref{eq:GG}), $K_i,(i=1,2)$ are the modified Bessel functions
of order $i$. Eq.~(\ref{eq:Blz}) is only sensitive to the
value of $\lambda(x)$ for $x \la 0.1$ which roughly corresponds to
the freeze-out temperature, and the center-of-mass energy
$\sqrt{s} \geq 2m_\chi$, as a result the argument of $K_1$ in
Eq.\ (\ref{eq:GG}) $\sqrt{s}/xm_\chi \ga 20$ so that $K_1$ dies away
quickly with increasing $\sqrt{s}$, owing to the asymptotic behavior
of the Bessel function ($K_1(y) \sim \sqrt{\pi/2y}e^{-y}, y \gg 1$).
Therefore, replacing $w(s)$ in (\ref{eq:GG}) by its
series expansion around $\sqrt{s}=2m_\chi$ and the Bessel functions
by their asymptotic expansions for large argument,
the resulting series of integrals can be readily carried out
analytically leading to \cite{Watkins}
\begin{eqnarray}
\langle{\sigma v_{\mol}}\rangle &=&{1\over m^2_\chi}
[w-{3\over 2}(2w-w')x+{3\over 8}(16w-8w'+5w'')x^2+{\cal O}(x^3)]_
{s=4m^2_\chi} \nonumber\\
&\equiv &a+bx+cx^2+{\cal O}(x^3),
\label{eq:abc}
\end{eqnarray}
where primes denote derivatives with respect to $s/4m^2_\chi$
rather than $s$ itself.

Contrary to Eq.\ (\ref{eq:GG}) which inevitably
requires numerical integration,
the expansion such as Eq.\ (\ref{eq:abc}) or its
non-relativistic counterpart is simple to use and
gives overall fairly good results. In fact, most of the relic
abundance calculations were carried out using such thermal
average expansion \cite{Wein,EHNOS,Group1,Group2}.
Recently, however, the degree of accuracy of such expansions has
been reexamined, and now it is clear that
such treatment actually fails badly near the
$s$-channel resonances and/or new-channel thresholds,
basically because the expansion of $w(s)$
(or equally, cross section for this matter)
at $\sqrt{s}=2m_\chi$ becomes inappropriate
in those cases \cite{GG91,GS91}. In addition,
in Ref.\ \cite{LNY93} it is shown with explicit examples that,
when close to the $s$-channel resonances the first-order expansion
$a+bx$ is much better behaved than the second-order
expansion $a+bx+cx^2$, in fact latter renders the thermal
average factor negative right above the poles which doesn't
make sense. Of course, in this case, one should abandon the
expansions and use the exact integral expression (\ref{eq:GG})
to get reliable results. Later in subsection 2.4,
I will discuss a practical situation where a proper treatment
of the thermal average factor is needed and
present the results calculated with Eq.~(\ref{eq:GG}) and
with first-order expansion $a+bx$.

\vglue 0.2cm
{\elevenit\noindent 2.2 Organizing Model Parameters: Supergravity Models}
\vglue 0.1cm
So far, I have kept my discussion fairly model-independent,
obviously, to go further we need the information of specific
supersymmetric model. Primarily, the details of the particle
physics model enter the function $w(s)$, which is
related to the annihilation cross section of neutralino pair
into all kinematically accessible final states.
What makes the analysis of these cross section fairly complicated
is mainly the fact that supersymmetric models often contain many free
parameters which to certain degree are rather arbitrary.
For instance, in the minimal supersymmetric standard model
(MSSM) \cite{MSSM}, which happens to be the playing ground of
almost all previous studies on neutralino dark matter
\cite{Wein,EHNOS,Group1,Group2}, even if the
mixing between generations are neglected
and only the Yukawa terms for the third generation are kept,
there are still at least 21 free parameters, many of which
describe the soft supersymmetry breaking at low energies.
With so many parameters, any thorough analysis becomes hopeless.
Nevertheless, to make life easy, in all these earlier works,
several additional assumptions about these parameters are made,
notably, the GUTs relation among different gaugino masses,
and a common mass parameter for all sleptons and squarks.
This situation is hardly satisfactory. We know that supersymmetry
must be broken, however, just as we do not really
understand the gauge symmetry breaking mechanism, our knowledge
about supersymmetry breaking is even more limited.
The phenomenology of low-energy supersymmetry should
reflect how supersymmetry is broken, and therefore could
in principle provides us with some information on the
pattern of supersymmetry breaking. Since any such information
is precious, it would thus be very important if one can
also obtain constraints on the soft supersymmetry breaking
terms from the cosmology of neutralino dark matter.
Clearly, in the usual approach, particularly
due to the ad-hoc assumption about the masses of sleptons
and squarks, no ``fine structure'' of
supersymmetry breaking is left to be found. On the other hand,
as I mentioned earlier, to address such problem with all
the soft-breaking parameters being arbitrary simply is
not possible. One way out is provided by $N=1$ supergravity \cite{Nilles}.
Although it was realized long ago that local supersymmetry,
instead of global supersymmetry, should be the natural framework in
which to construct realistic low-energy supersymmetric models,
only very recently has the study of neutralino dark matter in
supergravity models attracted enough attention
\cite{EZ,Noji,LNY,LYN,ER92,Fiveh,DN93,Fivedm,Roberts}.

In supergravity models, in addition to
the observable sector, which contains quarks, leptons, Higgs bosons,
gauge bosons as well as their superpartners, and in case of a
grand unified model also the extra particles, there is also
a so-called hidden sector responsible for the spontaneous
breaking of $N=1$ local supersymmetry. At low energies, the
breaking of supergravity taking place in the hidden sector
transmits into the observable sector via gravitational
interaction, and therefore leads to a globally supersymmetric
effective theory with explicit soft-breaking terms.
One of the virtue of the supergravity models is that
the number of soft-breaking parameters are greatly reduced.
In fact, the supersymmetry breaking pattern now can be specified
mainly in terms of only three universal parameters at some
unification scale $M_U$: (I) the scalar mass $m_0$;
(II) the Majorana gaugino mass $m_{1/2}$;
and (III) the trilinear $A$ and bilinear $B$ scalar couplings.
Furthermore, in supergravity models the breaking of the electroweak
gauge symmetry can be realized as a consequence of the supersymmetry
breaking through radiative correction \cite{IR}, with
this radiative breaking mechanism implemented, other
model parameters can be further reduced. In the model
with minimal particle content same as that of MSSM,
up to the sign of Higgs mixing parameter $\mu$, one only
needs two more parameters to describe the whole model,
these can be chosen as the unknown top quark mass $m_t$ and
the ratio of Higgs vacuum expectation values
$\tan\beta = v_2/v_1$, and finally all the couplings
and superparticle masses are determined dynamically
via the relevant renormalization group equations (RGEs)
as functions of the above five parameters \cite{FiveAP}.

\vglue 0.2cm
{\elevenit\noindent 2.3 Supersymmetry Breaking and Dark Matter}
\vglue 0.1cm
I am now ready to describe the correlation between the neutralino
relic abundance $\Omega_\chi$ and the supersymmetry breaking
patterns, first found in Ref.\ \cite{LYN}, as one of the most
interesting features of the supergravity models we are considering.

The idea is very simple. Since now all the low-energy couplings
and superparticle masses entering $w(s)$ depend explicitly upon
soft-breaking parameters $m_0$, $m_{1/2}$ and $A$, the relic
abundance $\Omega_\chi$ itself becomes a function of these
parameters as well. The most crucial factors here are the
masses of the gauginos, squarks and sleptons at low energies,
which are determined as functions of soft-breaking parameters
through the renormalization equations.
First, the gaugino masses $(M_i, i=1,2,3)$
are given by $M_i=(\alpha_i/\alpha_U)m_{1/2}$,
where $\alpha_U=0.0409$ is the gauge coupling at the
unification scale $M_U \approx 10^{16}$ GeV (determined
from following input low-energy values: $\alpha_3=0.111$,
$\sin^2\theta=0.233$ and $\alpha=1/127.9$).
Second, neglecting the Yukawa coupling contributions,
the renormalization equations of the sfermion mass
can be solved exactly giving \cite{LKN}
\begin{eqnarray}
m^2_{\tilde f}&=&m^2_f+m^2_0+m^2_{1/2}c_{\tilde f}
-M^2_W{\tan^2\beta-1\over\tan^2\beta+1}
[(T_{3,f}-Q_f)\tan^2\theta_W+T_{3,f}] \nonumber \\
&=&m^2_f+(1.22M_2)^2(c_{\tilde f}+\xi^2_0)+D_{\tilde f}
\label{eq:sfm}
\end{eqnarray}
where $\xi_0\equiv m_0/m_{1/2}$, and the coefficients
$c_{\tilde f}$ are determined to be:
$c_{\tilde e_L,\tilde\mu_L,\tilde\nu}=0.514$,
$c_{\tilde e_R,\tilde\mu_R}=0.150$,
$c_{\tilde u_L,\tilde c_L}=c_{\tilde d_L,\tilde s_L}=6.134$,
$c_{\tilde u_R,\tilde c_R}=5.720$,
$c_{\tilde d_R,\tilde s_R}=5.670$.
The parameter $A$ enters the off-diagonal terms of the
third generation scalar masses. In Ref.\ \cite{LYN} only the
left-right mixing for stop quark masses is considered,
and the diagonal contributions for all the third generation
sfermions are approximated with the $c'$s given above
and Eq.~(\ref{eq:sfm}).
It is found that the major effect of $A$ is to restrict
the allowed parameter space, otherwise $A$ does not
change the neutralino abundance $\Omega_\chi$ significantly.
In what follows, I choose $A=m_0$.

In Figs.~1--3, the neutralino relic abundance $\Omega_\chi$
is shown in the $(\mu,M_2)$ plane for $\tan\beta=8$, $\mu>0$
($\mu<0$ case is similar)
\footnote{\ninerm\baselineskip=11pt Here the sign convention
of $\mu$ is opposite to that of Refs.\ \cite{LNY,LYN}.},
and three representative values of $\xi_0$: (a) $\xi_0=10.0$,
(b) $\xi_0=1.0$, (c) $\xi_0=0.1$. Here only the tree-level
Higgs masses are used, and the lightest CP-even Higgs mass
is taken to be $m_h=45$ GeV, different choice of this parameter
leads to similar results. I also take $h_0=0.5$, a favorite
choice of cosmologists \cite{Turner}, and divide the allowed
parameter space into three types of distinct regions:
(1) regions represented by stars corresponds to $\Omega_\chi > 1.0$,
which are excluded cosmologically;
(2) regions represented by vertical
crosses corresponds to $0.1 < \Omega_\chi < 1.0$, which
are cosmologically favored; (3) regions represented by
dots corresponds to $\Omega_\chi < 0.1$, where neutralinos
can not even account for enough dark matter in Galactic halos.

In these figures, the shape of the allowed parameter space
in each case was determined by several constraints.
The direct experimental constraints used are: (1) The LEP lower
bound on the chargino mass $m_{\chi^\pm} > 45$ GeV, and on the
slepton mass $m_{\tilde l} > 43$ GeV;
(2) The CDF lower bound on the gluino mass $m_{\tilde g} > 150$ GeV,
which translates into $M_2 > 45$ GeV, and on the squark mass
$m_{\tilde q} > 100$ GeV. In addition, the assumption about neutralino
being the LSP imposes important consistency constraints on the
allowed parameter space as well. For example, in Fig.~3 a portion
of the upper right corner is excluded since there right-handed
sleptons become lighter than the neutralino. Also, in all these
figures, the allowed minimal value of $M_2$ is in fact bigger
than that required by the gluino mass lower bound; and when
$\xi_0$ decreases, this lower bound on $M_2$ increases. It
is because that we do not want the sneutrinos to be lighter
than the neutralino. Eq.~(\ref{eq:sfm}) clearly shows that
this constraint becomes stronger for small value of $\xi_0$.
Finally, in the lower right corner of these figures,
there is a triangle region where the left-right mixing term
for stop quark masses could drive the ${\tilde t}_1$ mass below
either the CDF bound or $m_\chi$, and thus it should be excluded.
This effect is only barely visible in Fig.~2, but it becomes
more pronounced for small values of $\tan\beta$.

As in the case of MSSM \cite{Group2}, Figs.~1--3 once again
exhibits the fact that the neutralinos relevant to cosmology should
either contain a dominant higgsino component or a
dominant bino component. In addition, two very interesting
new features emerge in Figs.~1--3, due to the supergravity squark
and slepton relation (\ref{eq:sfm}). First,
in the pure higgsino region (upper left portion of
the $(\mu, M_2)$ plane), we see that the relic abundance
$\Omega_\chi$ does not change with the different values
of $\xi_0$. This is because there the contributions to the
annihilation cross section due to the exchange of
sfermion have already been considerably suppressed,
so the variation of squark and slepton masses
with supersymmetry breaking patterns essentially has no effect
on $\Omega_\chi$ for nearly pure higgsinos.
Independent of how supersymmetry is broken,
pure higgsino dark matter candidate will
have a mass of roughly the order of $M_W$, but there
the gluino would be heavier than about $1.5$ TeV.
However, if one insists that $m_{\tilde g} \la 1$ TeV to insure
the fine-tuning of the parameters occurs at two-orders-of-magnitude
or less \cite{FiveAP}, this possibility of nearly pure
higgsino dark matter would be eliminated.
Second, in the pure bino region (lower right portion of the
$(\mu, M_2)$ plane), however, $\Omega_\chi$ changes
with $\xi_0$ significantly, mainly because the sfermion
exchange contribution is dominant in this region. From Eq.~(\ref{eq:sfm})
we see that large value of $\xi_0$ implies all sfermions are heavy, so
this contribution is suppressed, which leads to a large
value of $\Omega_\chi$. When $\xi_0$ decreases, the
annihilation due to sfermion exchange becomes increasingly
efficient, and therefore $\Omega_\chi$ reduces. Note that
$m_{\tilde f}$ now vary throughout the $(\mu, M_2)$ plane
due to their $M_2$ dependence, and in the case of stop
quark ${\tilde t}_{1,2}$ their $\mu$ dependence as well.

The above discussion is mainly concerned with the impact of
the supergravity relation Eq.~(\ref{eq:sfm}) for the squark
and slepton masses on the neutralino relic abundance $\Omega_\chi$.
In this analysis, the radiative electroweak gauge symmetry
breaking \cite{IR} and the radiative corrections to Higgs
masses \cite{HRC} were not considered. These two issues also
have important consequences on the cosmology of neutralino
dark matter \cite{DN93,Fivedm,Roberts}. The first new feature
is that, after enforcing the requirement of radiative electroweak
symmetry breaking, the allowed parameter space is considerably
reduced \cite{Fivedm}. In the usual $(\mu, M_2)$ plane, the
two coordinates can no longer vary independently. For example,
for $\xi_0 \la 1.0$, this correlation between $\mu$ and $M_2$
could eliminate a rather large triangle region below the
diagonal which are otherwise allowed, since $\mu \la m_{1/2}=1.22M_2$.
In addition, since the one-loop radiative corrected Higgs masses
vary continuously as one moves around the $(\mu, M_2)$ plane,
and they are normally bigger than their tree-level values,
the overall effect is a suppression of the relevant annihilation
rate and hence an enhancement of the relic abundance
$\Omega_\chi$ relative to that shown in Figs.1--3, as long as
the regions in comparison now are still allowed. Notice that
pure binos only couple to Higgs very weakly, so such an
enhancement of $\Omega_\chi$ mainly occurs in the ``mixed''
regions. This makes the ``mixed'' neutralino a possible
candidate to be the major component of the Galactic halo.
Of course, the qualitative feature of the novel correlation
between the neutralino relic abundance $\Omega_\chi$ and
the supersymmetry breaking patterns, as demonstrated in
Figs.1--3, remains the same. In fact, the prospects for neutralino
dark matter depend most strongly on the parameter
$\xi_0$ \cite{LYN,Fivedm}, which can be summarized as follows:
(1) For $\xi_0 \sim 1.0$ there is a wide range of the other
parameters such that $\Omega_\chi \sim 1.0$;
(2) For $\xi_0 \ll 1.0$ the relic abundance normally is
too small, but it is possible in this case that some
``mixed'' neutralinos may still be able to account for
the dark matter in the Galactic halo;
(3) For $\xi_0 \gg 1.0$ the relic abundance is almost always
much too large in conflict with current cosmological
observations, except the accidental circumstances where
the relic abundance could be locally diluted due to
the presence of resonances and thresholds in the
annihilation cross section.

\vglue 0.2cm
{\elevenit \noindent 2.4. The Minimal $SU(5)$ Supergravity Models}
\vglue 0.1cm
In the supergravity models, one of the crucial assumptions is that an
unification of gauge couplings and mass parameters takes place at
some high-energy scale $M_U$ not far from the Planck scale.
One simple way to realize this is to invoke a grand unification type of
symmetry. \footnote{\ninerm\baselineskip=11pt However, in the context
of superstring theories such unification arises naturally even in
the absence of a grand unification group.} Up to now, my discussion
of the correlation between the neutralino relic abundance $\Omega_\chi$
and the soft supersymmetry breaking patterns has not relied on the
details of any specific grand unification models. Now, I would like to
consider the implications of this generic result in the minimal $SU(5)$
supergravity model \cite{SUfive}.

At low-energies, the minimal $SU(5)$ supergravity model consists of
the normal light particles of the MSSM. Again, the masses and couplings
of these light particles are completely specified in terms of a few
parameters as described in subsection 2.2. The new feature of
this model (as in all grand unified models) is the existence
of the additional heavy degrees of freedom arisen around
$M_U \sim 10^{16}$ GeV, which would lead to new phenomena such as
proton decay \cite{Gla}. In the minimal $SU(5)$ supergravity model,
due to its supersymmetric nature, the usual proton decays through
dimension-six operators mediated by exchange of either heavy gauge bosons
or heavy triplet Higgs bosons, which plague the ordinary
non-supersymmetric $SU(5)$ model \cite{Gla}, are strongly suppressed
and place no danger as compared with the experimental limit.
However, the new proton decays through the genuinely supersymmetric
dimension-five operators \cite{pd5} mediated by exchange of heavy
higgsinos are not very strongly suppressed. Note, for the
dimension-five-induced amplitude the suppression factor is
$1/(M_UM_W)$, while for the normal dimension-six amplitude
it is $1/M^2_U$. These dimension-five-induced processes could
lead to rather dangerous proton decays, typically in the modes
$p \rightarrow {\bar \nu}_{\mu,\tau}K^+$ \cite{ENR}.

Recently, in light of the current experimental limit
$\tau_{p \rightarrow {\bar \nu}K^+} > 10^{32}$ yr,
detailed analysis of the dimension-five-induced proton decay
constraints have been carried out \cite{Dick,MATS,LNP}
in the minimal $SU(5)$ supergravity model.
Assuming the exchanged triplet higgsino mass to be bounded above
by $M_{{\tilde H}_3} < 3M_U$ \cite{Dick}, it is found that the
proton decay constraint is rather restrictive. In particular,
as for the soft supersymmetry breaking patterns, the proton deacy
favors a large value of $\xi_0$ \cite{Dick,LNP}. Again, this result
can be easily understood in view of Eq.~(\ref{eq:sfm}). Since
squarks and sleptons appear in the loops of the box diagrams
responsible for the dimension-five-induced proton decays,
large value of $\xi_0$ implies that such processes are
suppressed. From the discussion in previous subsection, however,
we see that the cosmology of neutralino dark matter, on the
other hand, disfavors large value of $\xi_0$. Therefore, a
delicate balance have to be attained in order to satisfy
these two constraints simultaneously in the minimal $SU(5)$
supergravity model.

The conflict between these two types of constraints in the minimal
$SU(5)$ supergravity model was first pointed out in Ref.\ \cite{LNZ},
which has since spurred further investigations on this
subject \cite{LNP,ANdm,LNY93}.
Considering the combined effect of these two constraints,
it is found that the allowed parameter space of the minimal
$SU(5)$ supergravity model is dramatically reduced (see below).
However, there is still a region in the parameter space
where both constraints are simultaneously satisfied.
It turns out that, in this region the neutralino $\chi$ is
actually very close to the lightest CP-even Higgs ($h$) and/or
$Z$-boson resonances, {\iitt i.e.},
$m_\chi \approx {1\over 2} m_{h,Z}$.
This fact has cast doubts \cite{ANdm} on the accuracy of the
results obtained in Refs.\ \cite{LNZ,LNP},
since there the first-order expansion of form (\ref{eq:abc})
has been used to approximate the thermal average factor,
which in general is not a valid approach (see subsection 2.1).
Two groups \cite{ANdm,LNY93} have recently reassessed this problem
by treating the thermal average factor properly.

In Ref.\ \cite{LNY93}, an extensive search of the parameter
space of the minimal $SU(5)$ supergravity model has performed.
In practice, a data set of the five-dimensional parameter space
is generated first, which gives adequate radiative electroweak
gauge symmetry breaking and satisfies all known current
phenomenological constraints as described in Ref.\ \cite{FiveAP}.
Then, the proton decay constraint is used to further reduce the
allowed points in the data set. In this step, the effects of
two-loop gauge coupling unification and light supersymmetric
thresholds have also been included \cite{LNPZ}. Finally,
for all the remaining points in parameter space, which have
$\tan\beta = 1.5,1.75,2.0$, the neutralino relic abundance
$\Omega_\chi$ is computed in two different approaches:
(a) the thermal average factor is treated accurately
using Eq.~(\ref{eq:GG}); (b) the thermal average factor is
treated approximately using first-order expansion of
form Eq.~(\ref{eq:abc}). The results are shown in
Fig.~4 and Fig.~5 respectively. These figures indicate
that indeed the cosmological constraint is very powerful:
all the points above the solid (dashed) line, allowed by
other constraints, are now excluded if $h_0=1.0$ ($h_0=0.5$).
This result does not change even if the exact thermal average
procedure is followed. From these two figures, it is clear that
some shifts of the points are noticeable, and the actual
structure around the resonances in Fig.~4, in particular that
close to the $Z$-pole ($m_\chi \sim {1\over 2}M_Z$),
is broader, shallower and asymmetric relative to Fig.~5.
Moreover, the cosmologically allowed points in Fig.~4 are
increased relative to that in Fig.~5. However, the qualitative
distributions of the points in Fig.~4 and Fig.~5 are similar,
and the difference between two cases appears to be not significant.

\vglue 0.5cm
{\elevenbf \noindent 3. Indirect Search for Neutralinos with
``Neutrino Telescope''}
\vglue 0.4cm
In this section, I wish to briefly report on our recent work \cite{NT}
about the possibility of indirectly detecting neutralinos of some
supergravity models with the so-called ``Neutrino Telescope'',
based on the assumption that neutralinos in these models constitute
the Galactic halo. Here, I will only present the result for the
minimal $SU(5)$ supergravity model, for detailed discussion see
Ref.\ \cite{NT}.
This subject has a rich history \cite{SEoo},
more recent work in the context of MSSM can be find
in Refs.\ \cite{RS88,GR89,GGR91,K91,Hal,KEK,Boti}.

If neutralinos populate the Galactic halo, some of them could be
trapped \cite{PS85,Gould} when traveling through the Sun or
Earth, after losing substantial amount of energy during their
interactions with nuclei. The captured neutralinos sink to the
core of the Sun or Earth, annihilate with one another therein,
then produce a shower of ordinary particles which further decay
or interact with solar or terrestrial material and finally lead
to energetic neutrinos as one sort of ultimate products.
The high-energy neutrinos coming from the Sun or the core of
the Earth can be detected in underground detectors, via either
(a) contained events, by looking for charged leptons within the
detector (useful for detecting both $\nu_e$ and $\nu_\mu$);
or (b) through-going events, by searching for upward-going
muons which are the products of the interaction of neutrinos
with rock below the detector (useful for detecting $\nu_\mu$).
I will only consider the neutrino-induced upward-going muon
events here, as it is most promising.

In the primary direct capture, a neutralino with mass $m_\chi$
passing through the Sun or Earth loses enough energy due to its
elastic scattering with nuclei, so that its velocity falls
below the escape velocity at one particular point inside the
Sun or Earth. The capture rate can be written as
\begin{equation}
C=\left({2\over 3\pi}\right)^{1/2}M_B
{\rho_\chi{\bar v}_\chi\over m_\chi}
\sum_i{{f_i\over m_i}\sigma_i X_i},
\label{eq:cp}
\end{equation}
where $M_B$ is the mass of the Sun or Earth, $\rho_\chi$ and
${\bar v}_\chi$ are the local neutralino density and velocity
in the halo respectively, $\sigma_i$ is the elastic
scattering cross section of the neutralino with the nucleus
of element $i$, and $X_i$ is a kinematic factor which can
be obtained from Eq.~(A10) of the second paper in Ref.\ \cite{Gould}.
The annihilation process normally takes a time scale much shorter than
the age of the Sun or Earth to reach equilibrium with the capture
process, in this case the neutralino annihilation rate,
entering the high-energy neutrino flux, is just half of that given
in Eq.~(\ref{eq:cp}).

The determination of neutrino flux or final detection rate is
normally complicated, basically because the differential energy
spectra of neutrinos depend on various subsequent physical processes
that take place, see {\iitt e.g} Ref.\ \cite{K91}. The rather limited
current knowledge about these processes remains the major source of
the uncertainties in this type of analysis. Nevertheless,
some resonable approximations can be made to render this part
of the problem tractable. For example, in Ref.\ \cite{RS88} the
neutrino spectrum from injected quarks and leptons was
calculated by using the Lund Monte Carlo, the same procedure
then was followed and refined in Ref.\ \cite{K91}. It turns
out that, for the neutrino-induced upward-going muons events, the
detection rate can be approximated as \cite{RS88,K91}
\begin{equation}
\Gamma=\kappa_B({C\over {\rm sec}^{-1}})({m_\chi\over {\rm GeV}})^2
\sum_i{a_ib_i}
\sum_F{B_F{\langle Nz^2 \rangle}_{Fi}}\ {\rm m^{-2}yr^{-1}},
\label{eq:det}
\end{equation}
where $\kappa_B=1.27\times 10^{-29}$ ($=7.11\times 10^{-21}$)
for neutrinos from the Sun (Earth); the $i$-sum is over muon
neutrinos and anti-neutrinos, and $a_i=6.8$ $(3.1)$,
$b_i=0.51$ $(0.67)$ for neutrinos (anti-neutrinos); the $F$-sum
is over all annihilation final states that produce high-energy
neutrinos ($\tau{\bar \tau}$, $c{\bar c}$, $b{\bar b}$,
$t{\bar t}$, $WW$, $ZZ$, $hA$ and $HA$ in the minimal $SU(5)$
supergravity model.), each with branching ratio $B_F$, and
${\langle Nz^2 \rangle}_{Fi}$ is the second moment of the
spectrum of tpye-$i$ neutrino from final state $F$ scaled
by $m^2_\chi$.

In Fig.~6, the predicted detection rate of upward-going muon events
in the minimal $SU(5)$ supergravity model prediction is shown
as a function of $m_\chi$. The top and bottom row correspond to
the event rate resulting from the capture of halo neutralinos
by the Sun and Earth respectively. Also shown in Fig.~6 as
solid lines are the current Kamionkande $90\%$ C.L. upper limits \cite{CCLL}
of $6.6\times 10^{-14} {\rm cm}^{-2}{\rm sec}^{-1}$ (Sun)
and $4.0\times 10^{-14} {\rm cm}^{-2}{\rm sec}^{-1}$ (Earth).
{}From this figure, we see that the minimal $SU(5)$ supergravity
model is not further constrained by this type of underground
experiments. However, it is interesting to note,
if the experiment limits can be improved in the near future
by a factor of 100, then the region above the dashed lines
could be explored.

\vglue 0.5cm
{\elevenbf \noindent 4. Conclusions}
\vglue 0.4cm
The neutralino dark matter problem within the supergravity framework
is considered. In the supergravity models, all the particle masses
and couplings can be specified in terms of three universal
soft supersymmetry breaking parameters $m_0$, $m_{1/2}$ and
$A$ at the unification scale $M_U$, along with low-energy
parameters $m_t$ and $\tan\beta$. Also, the electroweak
gauge symmetry can be broken radiatively. As a result,
the neutralino relic abundance becomes strongly depends
upon the way supersymmetry is broken. If the soft supersymmetry
breaking seed is primarily a universal scalar mass ($\xi_0 \gg 1.0$),
then the relic abundance is almost always much too large, and
the cosmological constraint is very strong. The exception occurs
if resonances or thresholds are present. On the other hand, if the
soft supersymmetry breaking seed is a dominant universal gaugino
mass ($\xi_0 \ll 1.0$), the cosmological constraint is rather
weak. In the case of $\xi_0 \sim 1.0$ there is normally a
wide range of the other parameters where neutralinos can
provide a closure relic density $\Omega_\chi \sim 1.0$. In
the minimal $SU(5)$ supergravity model, since the
dimension-five-induced proton decay also imposes very
restrictive constraint which is somewhat in conflict with
the cosmological constraint, the parameter space of this
model is dramatically reduced, therefore, makes this model
more easily testable. Assuming neutralinos constitute the
dark matter in the Galactic halo, the upward-going muon
events in underground detectors provide yet anther probe to
explore supergravity models. However, in order to obtain
further useful constraints for the minimal $SU(5)$ supergravity
model, the current experimental limits would have to improve
by a factor of about 100 in the future.

\vglue 0.5cm
{\elevenbf \noindent 5. Acknowledgments \hfil}
\vglue 0.4cm
It's a pleasure to thank my collaborators R. Gandhi,
S. Kelley, J.L. Lopez, D.V. Nanopoulos, H. Pois and A. Zichichi,
for their contributions on the subjects covered this talk.
This work has been supported by a World-Laboratory Fellowship.
\vglue 0.5cm
{\elevenbf\noindent 6. References \hfil}
\vglue 0.4cm

\begin{figure}
\vspace{5in}
%\special{psfile=fig1_talk.ps angle=90 hscale=65 vscale=65 hoffset=450}
\caption{The neutralino relic abundance distribution
for $\tan\beta=8$, $h_0=0.5$, $\xi_0=10$. The meaning of the three
different symbols: (1) stars ($\Omega_\chi > 1.0$);
(2) crosses ($0.1 < \Omega_\chi < 1.0$);
(3) dots ($\Omega_\chi < 0.1$).}
\end{figure}

\begin{figure}
\vspace{5in}
%\special{psfile=fig2_talk.ps angle=90 hscale=65 vscale=65 hoffset=450}
\caption{Same as Fig.~1, but with $\xi_0=1.0$.}
\end{figure}

\begin{figure}
\vspace{5in}
%\special{psfile=fig3_talk.ps angle=90 hscale=65 vscale=65 hoffset=450}
\caption{Same as Fig.~1, but with $\xi_0=0.1$.}
\end{figure}

\begin{figure}
\vspace{6in}
%\special{psfile=fig4_talk.ps angle=90 hscale=62 vscale=62 hoffset=450}
\caption{The neutralino relic abundance in the minimal $SU(5)$
supergravity model as a function of $m_\chi$ calculated using
the exact thermal average procedure. The points above the solid
(dashed) line are excluded if $h_0=1.0$ ($h_0=0.5$).}
\end{figure}

\begin{figure}
\vspace{6in}
%\special{psfile=fig5_talk.ps angle=90 hscale=62 vscale=62 hoffset=450}
\caption{Same as Fig.~4, but the thermal average factor is approximated
with the first-order expansion.}
\end{figure}

\begin{figure}
\vspace{7in}
%\special{psfile=fig6_talk.ps angle=90 hscale=63 vscale=63 hoffset=450}
\caption{The detection rate for the neutrino-induced upwards-going
muon events in the minimal $SU(5)$ supergravity model
as a function of $m_\chi$. The top and bottom row show the event rate
resulting from the capture of halo neutralinos by the Sun and Earth
respectively. The solid lines in the figures are the
corresponding $90\%$ C.L. upper limits of Kamiokande. The dashed
lines indicate where such upper limits will be in the future
if the experiment capability is improved by a factor of 100.}
\end{figure}

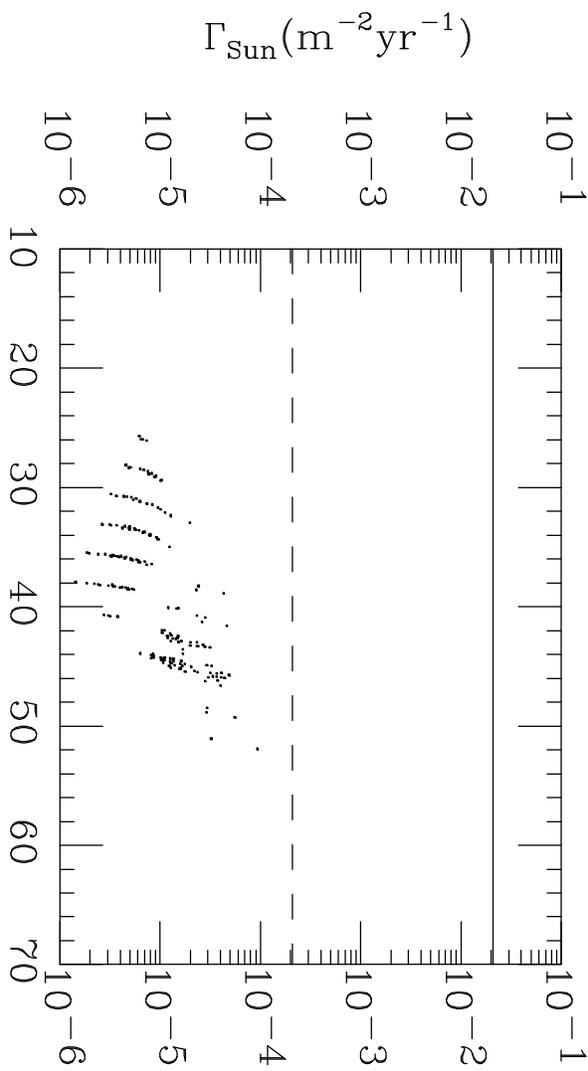
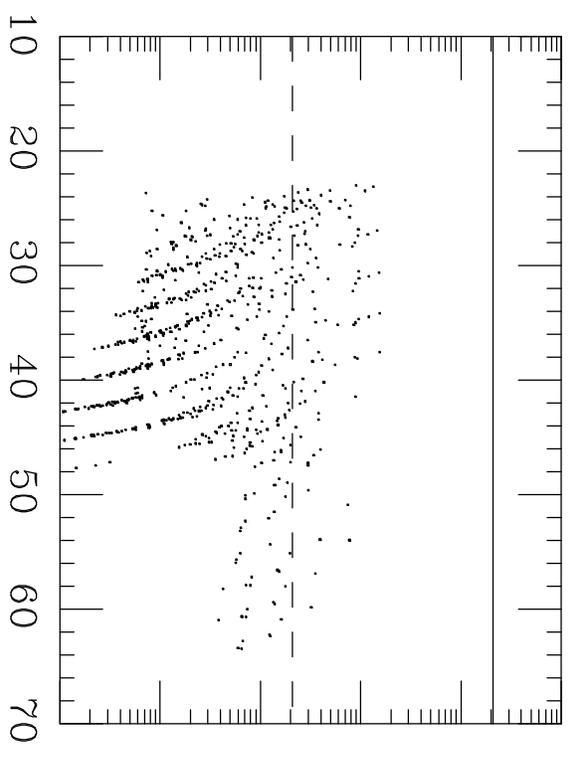
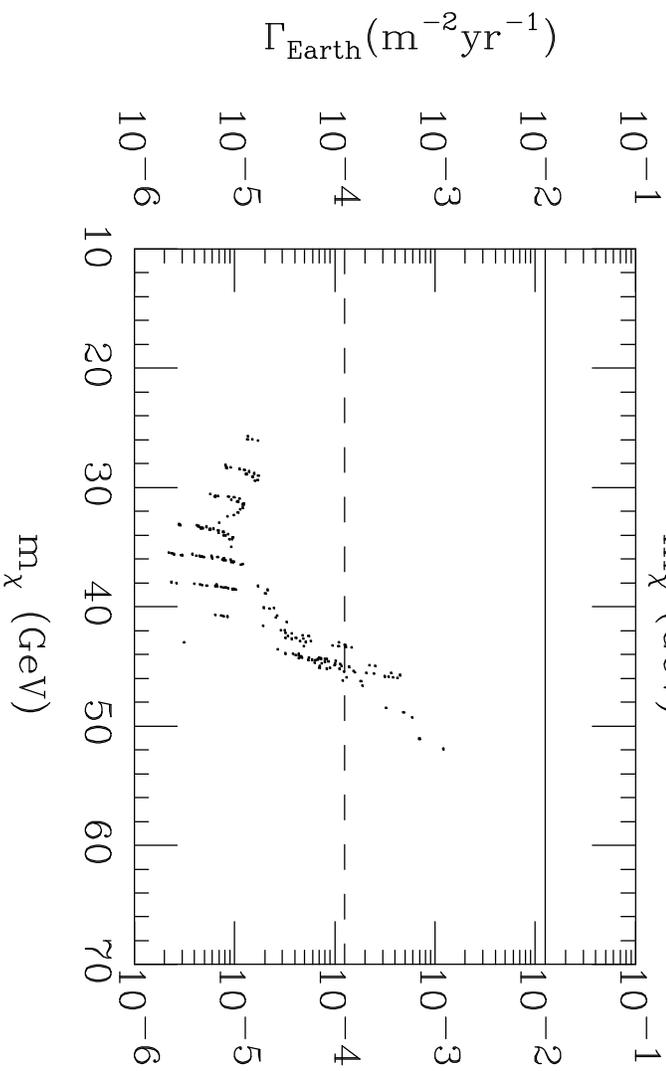
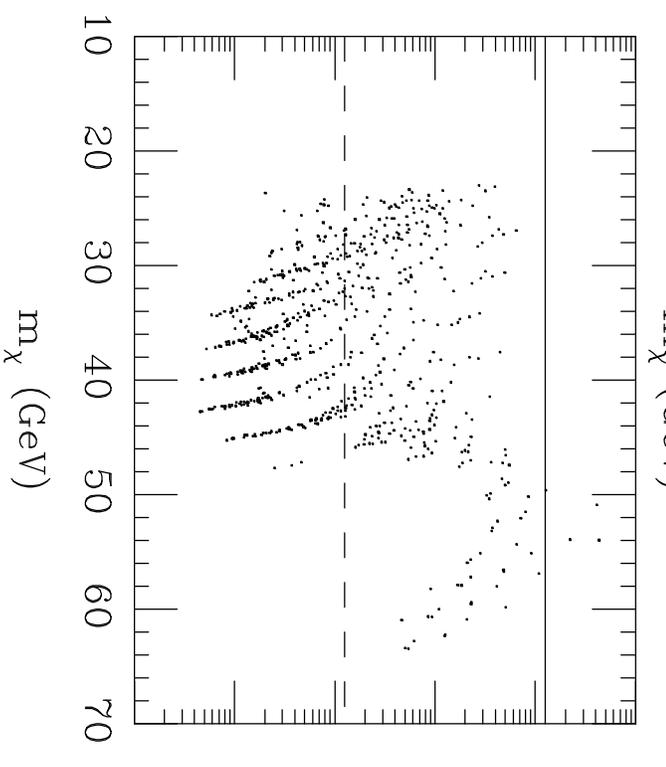
\end{document}